\documentclass[preprint]{iucr}

\usepackage{graphicx}
\usepackage[utf8]{inputenc}
\usepackage{url}
\usepackage{float}

     \paperprodcode{a000000}      
     \paperref{xx9999}            
     \papertype{FA}               

     \paperlang{english}          
     \journalcode{J}              
     \journalyr{2009}
     \journaliss{}
     \journalvol{}
     \journalfirstpage{000}
     \journallastpage{000}
     \journalreceived{0 XXXXXXX 0000}
     \journalaccepted{0 XXXXXXX 0000}
     \journalonline{0 XXXXXXX 0000}

\begin{document}                  



\title{Atomic pair distribution function analysis from the ARCS chopper spectrometer at the Spallation Neutron Source}

\author[a,b]{E.~S.}{Bo\v{z}in}
\author[a]{P.}{Juh\'{a}s}
\author[a]{W.}{Zhou}
\author[c]{M.~B.}{Stone}
\author[c]{D.~L.}{Abernathy}
\author[c]{A.}{Huq}
\cauthor[a,b]{S.~J.~L.}{Billinge}{sb2896@columbia.edu}{}


\aff[a]{Department of Applied Physics and Applied Mathematics, Columbia University, New York, New York, 10027, USA}


\aff[b]{Condensed Matter Physics and Materials Science Department, Brookhaven National Laboratory, Upton, New York, 11973, USA}


\aff[c]{Neutron Scattering Sciences Division, Oak Ridge National Laboratory, Oak Ridge, Tennessee, 37831, USA}

\maketitle                        

\begin{synopsis}
The first use of the ARCS chopper spectrometer at the Spallation Neutron Source at Oak Ridge National Laboratory
for powder diffraction based atomic pair distribution function (PDF) analysis is reported.
Its performance is benchmarked against the NPDF diffractometer at Los Alamos National Laboratory.
\end{synopsis}

\begin{abstract}
Neutron powder diffraction based atomic pair distribution functions (PDFs) are reported
from the new wide angular-range chopper spectrometer ARCS at the Spallation Neutron Source (SNS) at Oak Ridge
National Laboratory (ORNL). The spectrometer was run in white-beam mode with no Fermi chopper. The PDF
patterns of Ni and Al$_2$O$_3$ were refined using the PDFfit method and the results compared to data
collected at the NPDF diffractometer at Los Alamos National Laboratory. The resulting fits are of high
quality demonstrating that quantitatively reliable powder diffraction data can be obtained from ARCS when
operated in this configuration.
\end{abstract}



Neutron powder diffraction measurements can be carried out routinely using time-of-flight
(TOF) instruments at spallation sources such as the NPDF
at the Los Alamos Neutron Science
Center (LANSCE). Here we show that quantitatively reliable PDFs can also be obtained from the ARCS
chopper spectrometer at SNS when operated without a Fermi chopper.
The Fermi chopper mechanism at ARCS is mounted on a motorized
translation table, allowing one to easily switch between two installed Fermi chopper slit packages and
an open `white-beam' position.  This feature, together with the demonstration
that quantitatively reliable
structural information can be obtained from ARCS, means that it becomes straightforward to analyze structure \emph{and} dynamics from the same sample without dismounting it from the instrument.

No routines currently exist for a complete time-focussing of diffraction data from ARCS, and so the data
were histogrammed to produce the 1D powder diffraction pattern using a generic TOF to $d$-space conversion. This results in data with degraded
resolution, and with poorly defined peak shapes. Although the data could not be refined using the Rietveld method, we successfully obtained and refined atomic pair distribution functions (PDFs) to obtain quantitative structural information.

\begin{figure}
\includegraphics[width=0.65\textwidth,angle=270]{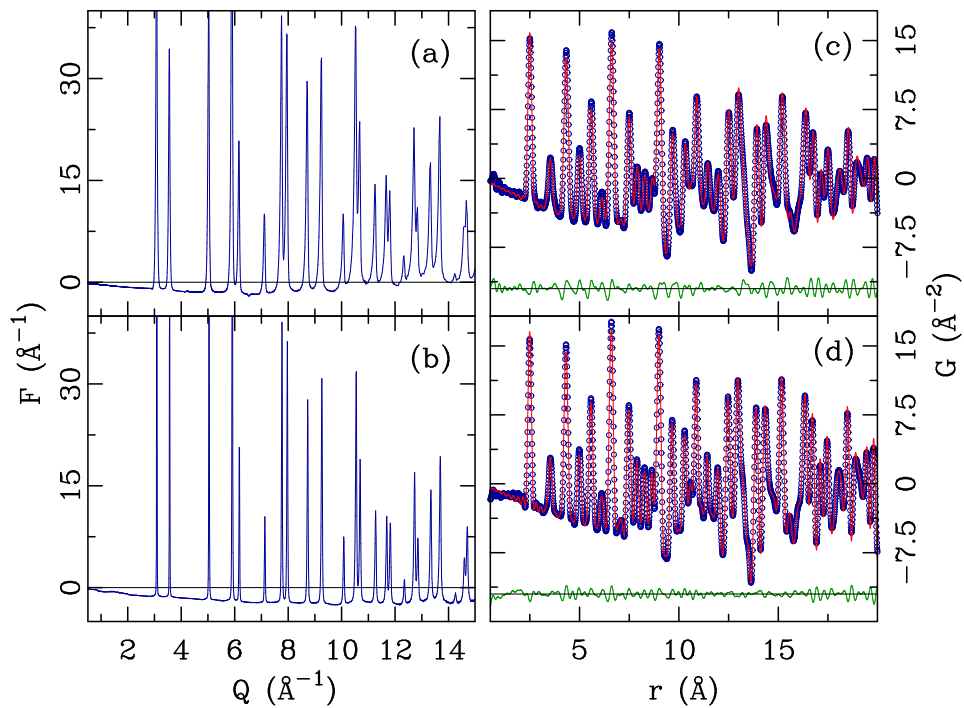}
\label{fig;FQNICKEL}
\caption{Total scattering function F(Q)=Q[S(Q)-1] of Ni from data collected at (a) ARCS and (b) NPDF.
Data were collected at room temperature for 30 minutes at both instruments. Corresponding experimental
PDFs (open symbols), PDFgui fits of the structural models (solid lines) and associated difference curves
(offset for clarity) for (c) ARCS and (d) NPDF.}
\end{figure}

Data were collected on two different commercially available
standard samples: Ni and Al$_2$O$_3$. For comparison, data from Ni
were obtained from the NPDF diffractometer at LANSCE. The samples were loose
powders of between 3~g and 8.3~g in weight sealed in extruded vanadium tubes.
A vanadium rod measurement was also performed to obtain incident spectrum information. A typical total scattering structure function, $F(Q)$ from ARCS is shown in Figure~\ref{fig;FQNICKEL}(a),
%
with the corresponding $F(Q)$ function from NPDF shown in Figure~\ref{fig;FQNICKEL}(b) for comparison. The resulting PDFs are shown in Fig.~\ref{fig;FQNICKEL}(c) and (d) with model fits, obtained using PDFgui~\cite{farro;jpcm07}, superimposed. The details of the PDF method are provided elsewhere~\cite{egami;b;utbp03}.

Simple visual comparison of the ARCS and the NPDF PDFs demonstrates that the ARCS data is of high
quality. Furthermore, all ARCS datasets could be successfully refined, including the more complicated Al$_2$O$_3$ structure. The quantitative results of the refinements are presented in Table~\ref{tab;PDFa}.
%
\begin{table}
\label{tab;PDFa}
\caption{Parameters from the PDF refinements of ARCS Ni and Al$_2$O$_3$ data at 300~K, compared to NPDF and the literature values~\cite{lewis;ac82}, respectively.  Ni (s.g. $Fm\overline 3m$) with Ni at (0,0,0). Al$_2$O$_3$ (s.g.
$R\overline 3c$) with Al at (0,0,$z$) and O at ($x$,0,0.25) positions.
}
\begin{tabular}{lllll}
\hline
Parameter&Ni(ARCS)&Ni(NPDF)&Al$_2$O$_3$(ARCS)&Al$_2$O$_3$ (lit.) \\
\hline
a (\AA)&3.5372(2)&3.5270(2)&4.7801(14)&4.7602(4)\\
c (\AA)&-&-&13.0502(16)&12.9933(17)\\
z$_{Al}$&-&-&0.3521(1)&0.35216(1)\\
x$_{O}$&-&-&0.3071(1)&0.30624(4)\\
U$_{iso}^{Ni}$ (\AA$^{2})$&0.0072(2)&0.0053(2)&-&-\\
U$_{iso}^{Al}$ (\AA$^{2})$&-&-&0.0055(2)&0.00347(3)\\
U$_{iso}^{O}$ (\AA$^{2})$&-&-&0.0061(3)&0.00419(3)\\
$\delta_{2}$ (\AA$^{2}$)&2.74(4)&2.77(2)&1.79(11)&-\\
scale&0.57(2)&1.03(3)&1.37(2)&-\\
Q$_{d}$ (\AA$^{-1}$)&0.046(11)&0.017(4)&0.046&-\\
r$_{W}$&0.108&0.067&0.109&-\\
\hline
\end{tabular}
\end{table}
As is evident from Tab.~\ref{tab;PDFa}, quantitative structural parameters can be refined from the ARCS
PDF with high accuracy. Interestingly, the peak width in the ARCS PDF
also appears to be larger than that for the NPDF data for the same Q$_{max}$ used.
This effect is known to appear when data have a $Q$-dependent peak broadening~\cite{toby;aca92} but appears to be particularly marked in the ARCS data.  If it is not accounted for in the model, this results in an overestimate for atomic displacement parameters from ARCS, as is evident in Tab.~\ref{tab;PDFa}. Notably, the fit gives a very good value for
the agreement factor r$_W$, that is a standard qualitative goodness of fit
measure~\cite{egami;b;utbp03}.


\ack{Acknowledgements}
Work in the Billinge-group was supported by by the Office of Basic Energy Sciences, U. S. Department
of Energy (BES-DOE) under contract no. DE-AC02-98CH10886.
The research at ORNL's SNS was sponsored by the Scientific User Facilities Division, BES-DOE.




\end{document}